\documentstyle[psfig]{laa}

\newcommand{\mdot}{\mbox{$\dot{M}$}}

\begin{document}

\thesaurus{08.16.4, 08.13.2, 13.09.6, 09.16.1, 09.04.1}
\title{Crystalline silicates in Planetary Nebulae with 
[WC] central stars\thanks{
Based on observations with ISO, an ESA project with instruments        
funded by ESA Member States (especially the PI countries: France
Germany, the Netherlands and the United Kingdom) and with the
participation of ISAS and NASA}}
\author{L.B.F.M. Waters\inst{1,2}, D.A. Beintema\inst{2}, 
A.A. Zijlstra\inst{3}, A. de Koter\inst{1}, 
F.J. Molster\inst{1}, J. Bouwman\inst{1}, 
T. de Jong\inst{2,1}, S.R. Pottasch\inst{4}, 
and Th. de Graauw\inst{2}}

\offprints{L.B.F.M. Waters (Amsterdam address)}
\institute{Astronomical Institute 'Anton Pannekoek', University of
Amsterdam, Kruislaan 403, NL-1098 SJ Amsterdam, the Netherlands 
\and
SRON Laboratory for Space Research Groningen, P.O. Box 800,
NL-9700 AV Groningen, The Netherlands 
\and
Department of Physics, UMIST, P.O. Box 88, Manchester M60 1QD,
U.K.
\and
Kapteyn Institute, University of Groningen, P.O. Box 800, NL-9700~AV
Groningen, The Netherlands}

\date{received date; accepted date}

\maketitle
\markboth{O-rich dust around [WC] stars}{L.B.F.M. Waters et al.}

\begin{abstract}

We present ISO-SWS spectroscopy of the cool dusty envelopes surrounding two
Planetary Nebulae with [WC] central stars, BD+30~3639
and He~2-113. The $\lambda$~$<$~15~$\mu$m region is dominated by a
rising continuum with prominent emission from C-rich dust (PAHs), 
while the long
wavelength part shows narrow solid state features from crystalline
silicates. This demonstrates that the chemical composition of both stars
changed very recently (less than 1000 years ago). The most likely
explanation is a thermal pulse at the very end of the AGB or shortly
after the AGB. The H-rich nature of the C-rich dust suggests that the
change to C-rich chemistry did not remove all H. The
present-day H-poor [WC] nature of the central star may be due to 
extensive mass loss and mixing following the late thermal pulse. 

\end{abstract}

\section{Introduction}

Planetary Nebulae (PNe) are the ionized and photodissociated 
remnants of extensive mass loss 
which the central star experienced when it was a cool 
Asymptotic Giant Branch (AGB) star. PNe therefore provide insight in the
mass loss on the AGB and beyond.
An interesting class of PNe are those with a Wolf-Rayet ([WC]) central 
star. These central stars have strongly enhanced C and He, but little 
or no H in their atmosphere, and are the low mass counterparts to 
the population~I Wolf-Rayet stars. About 50 [WC] stars are known
(Gorny \& Stasinska 1995). The properties of these nebulae do not differ
significantly from nebulae with 'normal' central stars
(Pottasch 1996). 

The formation of H-poor central stars is somewhat of 
a puzzle, but is probably related to a thermal
pulse either at the very end of the AGB or young PN phase (in this paper 
referred to as PAGB pulse; Zijlstra et al. 1991), or when the 
star is already on the cooling track (very late thermal pulse; Iben
1984). The H-rich layers may be removed due to efficient mixing
and subsequent nuclear burning, or by extensive mass loss, exposing
processed layers to the surface.
Examples of objects that may have recently experienced a late thermal pulse
are FG~SGe and Sakurai's object (e.g. Paczy\'nski 1970; Nakano et al 1996;
Duerbeck et al. 1996; Bl\"ocker \& Sch\"onberner 1997). 
It is not clear that these stars will develop [WC] type spectra. 

IR spectroscopy of the dust ejected by [WC] central stars
shows the well-known family of IR emission bands, usually identified as
due to Polycyclic Aromatic Hydrocarbons (PAHs) (e.g. Cohen et al. 1986).
This confirms the C-rich nature of the most recent mass loss episode.
In this {\em Letter}, we present new infrared spectra taken with the
Short Wavelength Spectrometer (SWS) on board of the Infrared Space
Observatory (ISO) of two well-studied PNe with [WC] central stars, 
BD+30~3639 and He~2-113. Both nebulae have associated molecular 
gas (Taylor et al. 1990; Bachiller et al. 1991; Gussie \& Taylor
1995). In Sect.~2, we discuss the observations and data reduction,
Sect.~3 gives an inventory of the solid state features (both C-rich and
O-rich). In Sect.~4 we present a first attempt to model the dust
spectrum of BD+30~3639, and we discuss these results in the context of 
post-AGB evolution and formation scenarios for [WC] central stars of PNe.

\begin{figure}
\psfig{figure=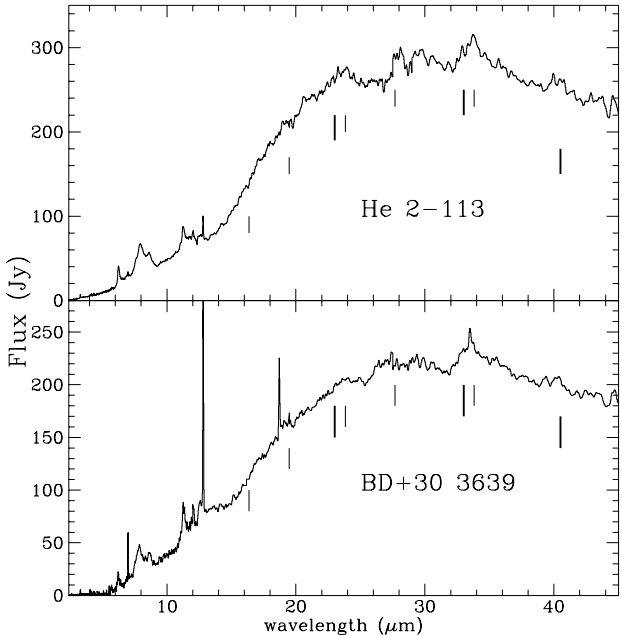,width=10cm}
\caption[spectra]{2.4-45 $\mu$m ISO-SWS spectra of He~2-113
and BD+30~3639, showing the PAH emission features at $\lambda$ $<$ 13 $\mu$m 
and the crystalline silicates at $\lambda$ $>$ 20 $\mu$m. Tickmarks indicate
olivine bands (short thin lines; Dorschner, private communication) and 
pyroxene (long thick lines; Koike et al. 1993). Emission from [Ne~II] 
at 12.8 $\mu$m is seen in both nebulae, and from [Ar~II] at 6.99 $\mu$m, 
[S~III] at 18.71 $\mu$m, and [Si~II] at 34.8 $\mu$m in BD+30~3639.}
\end{figure}

\section{The Observations}

Full scans (AOT01 speed 1 and 2, $\lambda$ 2.4-45 $\mu$m) of BD+30~3639
([WC9]) and He~2-113 ([WC11]) were obtained using the SWS 
(de Graauw et al. 1996) on board 
of ISO (Kessler et all. 1996), as part of a guaranteed time programme on
spectroscopy of PNe (p.i. D.A. Beintema). The spectra were taken 
on Nov. 6, 1996
(BD+30~3639) and February 4, 1996 (He~2-113), and were reduced using
version 5.3 of the SWS off-line analysis pipeline. All detector signals
were inspected for spurious jumps and glitches, and in two
cases removed manually. The spectra of the 12
individual detectors in each SWS AOT band were sigma-clipped, 
averaged and re-binned  to a
uniform resolution of 250. Finally, the different AOT bands 
were joined together to form a continuous spectrum. 
For He 2-113 we did not need to adjust individual
AOT bands, but for BD+30 3639 we reduced the fluxes measured in bands 3E and 4
(wavelengths of 27.5 microns and longer) by 17 per cent.  This is within the
absolute flux calibration uncertainties (Schaeidt et al.  1997).  Jumps in flux
can occur at 27.5 and 29.0 microns, where the effective aperture changes from
14*27 to 20*27 and 20*33 arcsec).  The jump seen in BD+30 3639 could be
caused by the finite extent of the nebula (10 arc sec or less in the optical,
Harrington et al.  1997); Cox et al. (1997, in preparation) find that 
the extent of the neutral envelope is twice that of the ionized nebula.
Flux jumps can also be due to a small pointing error.  The good agreement
with the IRAS-LRS spectrum (not shown) excludes a serious pointing error. 
The resulting spectra
are plotted in Fig.~1, and in Fig.~2 we plot continuum subtracted 
spectra for both objects. The location of the continuum was estimated by
eye. This procedure may underestimate the contribution from weak, broad
features that are not easily distinguishable from the smooth underlying
continuum. However narrow (width less than a few $\mu$m) features are
well preserved. 

\section{Solid state features}

Here we discuss the solid state features in our SWS spectra. 
A full description of the atomic and ionic emission lines 
will be given elsewhere.

The short wavelength part of the spectra of both stars is dominated by
prominent emission from the well-known family of IR emission bands,
usually attributed to Polycyclic Aromatic Hydrocarbons (PAHs)
(e.g. L\'eger \& Puget 1984). PAHs in
BD~+30~3639 were previously reported by e.g. Witteborn et al. (1989),
Allamandola et al. (1989), and Cohen et al. (1986, 1989).
The overall shape of the PAH bands is similar in both objects (Fig.~2).
The plateau at 11-14 $\mu$m is very prominent in BD+30~3639, but does
not extend as far to the red as in IRAS21282+5050 (Beintema et al.1996;
Molster et al. 1996).
This plateau is attributed to a blend of many bands due to deformation
modes of large PAH molecules. As in IRAS21282+5050, we find prominent 
12.0 and 12.7 $\mu$m features. These latter bands may be attributed to C-H
bending modes in PAH molecules with 2 or 3 adjacent H atoms attached
(Allamandola et al. 1986). This evidence of H in the
C-rich dust shell is consistent with a recent determination of the photospheric
H abundance in He 2-113 (7 per cent, Leuenhagen \& Hamann 1997), although a
quantitative comparison of the present-day photospheric 
H abundance and the H content of the C-rich dust shell is difficult to make. 
The PAHs in BD+30~3639 suggest that also in that star H was present 
in the C-rich phase. 

In both objects, the spectrum at $\lambda$ $>$ 20~$\mu$m shows 
weak, and narrow solid state features, whose strength is 
of the order of 
10 per cent of the local continuum (Fig.~2). These solid state features are
observed in the spectra of {\em oxygen-rich} (proto-) planetary nebulae
and red AGB stars (Waters et al. 1996; Justtanont et al. 1996).
Following Waters et al. (1996) we identify the features at 19.8, 23.5,
27.5 and 33.8 $\mu$m with crystalline olivine (Koike et al. 1993;
Dorschner, private communication). The peak at 40.5 $\mu$m
is due to crystalline pyroxenes (Koike et al. 1993; Dorschner,
private communication; J\"ager et al. 1994). The pyroxenes also
contribute to the peaks at 23 $\mu$m and 33 $\mu$m. The band
strength ratio of the 23.5 and 33.8 $\mu$m bands of BD+30~3639 and
He~2-113 (0.5 and 1 respectively) suggests that the crystalline dust in
He~2-113 is warmer than that in BD+30~3639 (assuming similar dust
composition). This is consistent with the slope of the continuum. We
conclude that both [WC] stars experienced strong mass loss as an O-rich
star at the end of the AGB, and that the change to C-rich chemistry
immediately followed. The O-rich nature of the outer
layers of the BD+30~3639 nebula can explain the high albedo of the
scattering particles in the halo surrounding the ionized inner nebula
(Harrington et al. 1997).

\begin{figure}
\psfig{figure=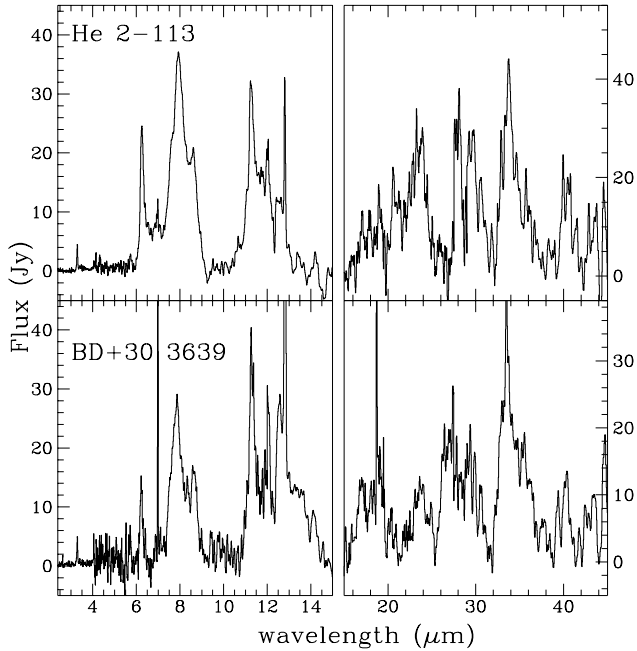,width=10cm}
\caption[pahs]{Continuum subtracted 2.4-45 $\mu$m spectra of He~2-113 (top)
and BD+30~3639. Notice the prominent UIR emission bands at 3.3, 6.2, 7.7, 
8.6, and 11.3 $\mu$m, often attributed to Polycyclic Aromatic
Hydrocarbons. At longer wavelengths emission from crystalline silicates
is found (olivines, pyroxenes).}
\end{figure}

\section{Discussion}

\subsection{A dust model} 

The presented model is calculated using the dust radiation transfer 
code {\sc modust} (de Koter et al. in preparation).
We have modeled the SWS spectrum of BD+30~3639 using a two component
dust model, of which the inner and outer shell connect. 
The inner part contains amorphous carbon grains and we have constrained 
the radial dimensions of this shell by requiring that it coincides with
the region from which PAH emission is observed (Bernard et~al. 1994),
i.e. from about 1.9'' to 2.7''.
Adopting a distance of 2.6 kpc (Hajian \& Terzian 1994), this yields
an inner and outer radius of the C-rich shell of 7.3~10$^{16}$ and
10.6~10$^{16}$ cm respectively. 
The luminosity of BD+30~3639 was set to 22,000 L$_{\odot}$ and
its effective temperature to 34.2 kK. Both these values are derived
from the analysis of Siebenmorgen et~al (1994) after scaling
to our (larger) adopted distance. Note that T$_{\rm eff}$ is rather
uncertain as estimates of this value range from 30 kK (Siebenmorgen
et~al. 1994) to 47 kK (Leuenhagen et~al. 1996).
The stellar luminosity yields a current mass $M$ = 0.89 $M_{\odot}$
(Paczy\'nski 1970), neglecting the possible effects of hot-bottom
burning.

Radiative transfer in the inner dust region is properly taken into account
and the emerging spectrum is used to irradiate the outer shell. The 
outer shell is assumed to be optically thin. Feed-back from the outer 
to the inner shell has not been taken into account, which is a valid
assumption as the O-rich outer shell is optically thin at all wavelengths.
The outer shell contains a mixture of amorphous and crystalline 
silicates. We use olivine (Dorschner, private communication), 
ortho-pyroxene and clino-pyroxene (Koike, private communication), and 
amorphous silicates (Draine \& Lee 1984) with 12, 1, 2 and 85 per 
cent abundance respectively (by mass), and a grain size radius of 
0.2 $\mu$m (0.25 $\mu$m for the amorphous silicates). The outer radius 
of the O-rich shell is 4.4~10$^{17}$ cm, but is poorly constrained by the 
model.

The resulting model fit is shown in Figure~3. Note that our solid state 
model does not yet include all structures seen in the spectrum; this 
will improve with better laboratory data. We find mass-loss rates for
the C-rich and O-rich shells of 2~10$^{-5}$ and 8~10$^{-5}$ M$_{\odot}$/yr
respectively, assuming a gas over dust ratio of 100 in both shells.
The errors in the mass loss are at least a factor of two. Reasons for
this large uncertainty are, among others: 
{\em (i)}   the poorly known gas/dust ratio,
{\em (ii)}  the sensitivity of \mdot\ to the grain size distribution
            function (we adopt grain sizes from 0.008 to 0.1 micron for
            the amorphous carbon in the inner shell and constant values
            in the outer shell),
{\em (iii)} the sensitivity of the emission of the O-rich material to 
            the internal nebular extinction from the C-rich dust (and 
            the ionized gas).

The uncertainty in mass loss also reflects on the accuracy of the
derived masses, which are $\sim$ 0.01 and $\sim$ 0.37 M$_{\odot}$ for 
the inner and outer shell respectively.
The O-rich shell mass is even more uncertain, as it depends also on the
poorly constrained outer boundary radius.
The inner radius of the O-rich shell corresponds to a dynamical age
of 1050 yr, assuming an expansion velocity of 22 km/sec (Acker et al. 
1992). Taking the O-rich mass at face value, this would imply
that this shell has been ejected over a time period of approximately
5000 yr. This is considerably shorter than the typical inter-puls
period. The mass loss in the O-rich shell is characteristic for OH/IR
stars. It may be usefull to search for OH maser emission, as one expects
that OH emission remains significant until $\sim$ 1000 to 1500 yr
after the end of the AGB.

\begin{figure}
\psfig{figure=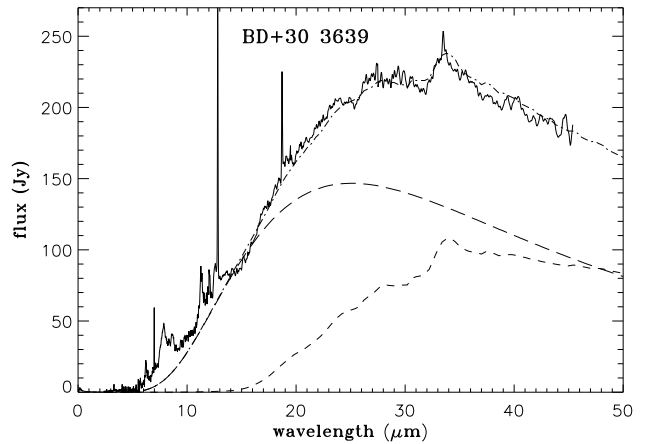,width=9.0cm}
\caption[model]{Model fit to the ISO-SWS spectrum of BD+30~3639. The
long dashed line is the C-rich dust shell, the short-dashed line
represents the O-rich detached shell; The dash-dotted line is the sum of
both components}
\end{figure}

\subsection{The evolution of [WC] stars}

The SWS spectra shown here demonstrate that both BD+30~3639 and He~2-113
were O-rich very recently. The age of the O-rich AGB 
remnant ($\sim$~1050~yrs in BD+30~3639, and possibly less for He~2-113) 
and the  mass of the C-rich envelope in BD+30~3639, suggest that the 
change to C-rich chemistry was 
triggered by a PAGB pulse and not by a very late thermal pulse.  The mass
contained in the C-rich dust shell of BD+30~3639
is of the same order of magnitude as the sum of the H and He layers 
in a young post-AGB star (Bl\"ocker et al. 1997). This suggests that
the PAGB pulse triggered the removal of the entire outer
layer of the star, exposing the He- and  C-rich inner layers. 
Evolutionary calculations of Bl\"ocker \&
Sch\"onberner (1997) show that in a PAGB thermal pulse,
no efficient mixing of the H-rich envelope into the 
He-burning layers is expected, and hence no [WC] star would form. Only
in the case of very late thermal pulses, when the central star is
already on the cooling track, mixing of H into the He
burning layers efficiently removes H, due to the strongly reduced 
envelope mass. Since the observations are not consistent with a very late
thermal pulse for these two nebulae, we suggest that a PAGB thermal pulse
is capable of mixing H and He-rich layers, i.e. contrary to the findings
of Bl\"ocker \& Sch\"onberner (1997). 

Recent calculations
of Herwig et al. (1997) suggest that a better treatment of convective 
mixing may result in intershell abundance patterns consistent with the
surface C, O, and He abundances of [WC] central stars 
(see also Leuenhagen \& Hamann 1997). The
latter authors show that late type [WC] stars have small amounts of H
and N, which may result from substantial mixing of H and He-rich layers.
It is likely that the thermal pulse which caused the change in chemistry
is responsible for this mixing. 

The production of [WC] central stars with O-rich outer shells 
obviously requires fine-tuning of the timing of the 
last thermal pulse, either at the very end of the AGB or shortly thereafter. 
Zijlstra et al. (1991) discovered OH maser emission from
IRAS07027-7934, demonstrating that the cool dust is O-rich. Barlow (1998)
shows SWS and LWS spectra of CPD-56~8032, also with O-rich dust. This implies
that O-rich cool dust is not unusual in nebulae with [WC] central stars. We
note that there is evidence that very late thermal pulses can 
also produce nebulae with [WC] central stars. 
Pollacco \& Hill (1994) claimed that the [WC11] star 17514$-$1555
(PN012.2+04.9) is surrounded by both a very compact nebula and an
extended, low density nebula, consistent with a very late thermal pulse. 
This is one of the few late-type [WC] stars which is relatively faint in 
the IRAS bands. It is unclear at present which mechanism (PAGB thermal pulse
with change in chemistry or very late thermal pulse) dominates the
formation rate of PNe with [WC] central stars.

\vspace{0.1cm}

\noindent{\bf Acknowledgements.} 
LBFMW and AdK gratefully acknowledges financial support from an NWO 'Pionier' 
grant. FJM acknowledges support from NWO grant 781-71-052. We thank
X. Tielens for discussions on PAHs, and H. Dorschner and C. Koike for 
providing laboratory data of dust species. We thank one of the referees, 
P. Cox, for communicating results on BD+30~3639 before publication.

\end{document}